\documentclass[%
 reprint,showkeys,
showpacs,preprintnumbers,
bibnotes,
prb,
]{revtex4-1}

\usepackage{graphicx}
\usepackage{dcolumn}
\usepackage{bm}

\begin{document}


\title{Metastability in the formation of Condon domaines}

\author{Leonid Bakaleinikov}

\affiliation{A. F. Ioffe Physico-Technical Institute, Russian Academy of Sciences, \\ St. Petersburg, 194021, Russian Federation}
\affiliation{ Department of Exact Sciences, Faculty of Natural Sciences, University of Haifa, Campus Oranim, Tivon 36006, Israel}

\author{Alex Gordon}
\affiliation{
 Department of Exact Sciences, Faculty of Natural Sciences, University of Haifa, Campus Oranim, Tivon 36006, Israel \\
}



\date{\today}

\begin{abstract}
 Metastability effects in the formation of Condon non-spin magnetic domains are considered. A possibility for the first-order phase transition occurrence in a three-dimensional electron gas is described in the case of two-frequency de-Haas-van Alphen magnetization oscillations originating from two extremal cross sections of the Fermi surface. The appearance of two additional domains is shown in the metastable region in aluminum. The phase diagram temperature-magnetic field exhibits the presence of second-order and first- order phase transitions in the two-frequency case. 
 
\end{abstract}

\pacs{71.10 Ca; 71.70 Di; 75.60.- d; 74.62 -- c; 75. 30. Kz;}
\keywords{Electron gas in quantizing magnetic fields; de Haas-van Alphen effect; diamagnetic phase transitions; Condon domains; metastability.}
\maketitle


\section{Introduction}
    Domain formation related to orbital magnetic moments of conduction electrons and their ``magnetic'' interaction \cite{b1} have been detected in silver, beryllium, white tin, indium, aluminum and lead by nuclear magnetic resonance (NMR) and muon spin-rotation spectroscopy (µSR) \cite{b2,b3,b4,b5,b6,b7,b8,b9}. These domains were observed in the de Haas-van Alphen (dHvA) oscillations of the orbital part of magnetization. The reason for the stratification of metallic samples into Condon non-spin domains is the instability of an electron gas called ``diamagnetic phase transition''\cite{b3}. This ordering occurs when the internal field in the sample is different from the external magnetic field (the Shoenberg effect) \cite{b1}. Properties of Condon domains in a two-dimensional electron gas have been considered in \cite{b11,b12,b5}. The direct observation of domains in silver has been achieved by Hall probes \cite{b7}. The hysteresis loop due to domains has been observed and investigated in beryllium \cite{b14}. Static, dynamic, size and tunneling effects close to phase transitions have been examined in \cite{b15,b16,b17}. In recent years there has been renewed interest in the problem of interactions between electrons leading to phase transitions in new measurements of magnetization oscillations, spectroscopy and Hall probes experiments \cite{b18} and astrophysics \cite{b19,b20}. Domain formation has been considered within the outer crust, deeper part of the internal crust and core of magnetars \cite{b21,b22}. 

So far, Condon domains and diamagnetic phase transitions have been considered for the simplest Fermi surface with one extremal cross section. A more realistic consideration of two frequencies dHvA oscillations caused by two extremal cross sections of the Fermi surface has not carried out for the examination of Condon domains formation. Up to now these domains have been thought to be formed as a result of a temperature second-order phase transition.  We will show that magnetization oscillations due to two extremal cross sections can lead to metastable phenomena in the domain formation.  

\section{Model}
    The oscillatory part of the thermodynamic potential density can be written by using  the Lifshitz-Kosevich formula in the `first harmonic approximation' \cite{b4}:
    
$$ \Omega ={\frac{1}{4\pi k^{2}} } \left[a\cos \left(b\right)+\frac{1}{2} a^{2} \sin ^{2} \left(b\right)\right]. \eqno (1) $$ 

Here $b=k(B-H)=k\left[h+4\pi M\right],$ {\it B} is the induction, {\it H} is the internal magnetic field, $k={2\pi F / H^{2}}  $, $F$ is the oscillation frequency, $h=H_{ex} -H$ is the small increment of the field ($H_{ex} $ the external magnetic field). In the `first harmonic approximation' the magnetization is found from the equation of state \cite{b1}:

 $$4\pi kM=a\sin \left[k\left(h+4\pi M\right)\right], \eqno (2) $$ 

where $a=4\pi kA=4\pi \left({\partial M / \partial B} \right)_{B=H} $ is the reduced amplitude of oscillations \cite{b1}, and {\it A }is the amplitude of the fundamental oscillation. According to \cite{b1}, the temperature and field dependence of the amplitude is 
$$a\left(T,T_{D} ,H\right)=a_{0} \left(H\right){\lambda T \exp \left[-\lambda \left(H\right)T_{D} \right]/  \sinh \left(\lambda T\right)}. \eqno (3)$$                                              

Here $\lambda \equiv {2\pi ^{2} k_{B} m_{c} c /( e\hbar H )} $, {\it mc }is the cyclotron mass, $k_{B} $ is the Boltzmann constant, {\it e} is the absolute value of the electron charge, {\it c }is the light velocity, {\it h} is the Planck constant, $T_{D} $ is the Dingle temperature. The limiting amplitude $a_{0} \left(H\right)$ is given by $a_{0} \left(H\right)\equiv a\left(\lambda T\to 0,0,H\right)=\left({H_{m} / H} \right)^{{3/ 2} } $ , where $H_{m} $ is the limiting field above which the diamagnetic phase transition does not occur at any temperature. For temperature-magnetic field phase diagrams giving the range of diamagnetic phase transitions one can use the equation for the phase transition temperature making the expression (3) equal to unity at the phase transition point \cite{b23}. 

      We present the thermodynamic potential density $\Omega $ as follows \cite{b1}:

 $$\Omega ={\frac{1}{4\pi k^{_{2} }} } \left[a\cos \left(kh+\mu \right)+{\frac{1}{2}} \mu ^{2} \right], \eqno (4)$$                                                                                     
where $\mu =4\pi kM$ is the dimensionless magnetization. The case of $h=0$ corresponds to the center of the dHvA period.

    Equations (1) and (4) present the approximated case of the simplest Fermi surface with one extremal cross section.  In many cases the presentation of the Fermi surface with two extremal cross sections corresponds to the nature of metal better than the approximated equations (1) and (4). We consider the two extremal cross sections case following Ref. \cite{b1}. Then the thermodynamic potential density is given by
$$\Omega ={\frac{1}{4\pi k^{_{2} }}  } \left[Ak\cos \left(\mu \right)+{\frac{1}{2}} \mu ^{2} \right]+ $$
$${\frac{1}{4\pi k'^{2}}  }
 \left[A'k'\cos \left(\mu '\right)+{\frac{1}{2}} \mu '^{2} \right], \eqno (5)$$                             
$\mu '=4\pi k'M$. Taking the amplitudes and frequencies of dHvA magnetization oscillations due to the two extremal cross sections of the Fermi surface $A$ and$A'$as $A'=\alpha A,k'=\beta k,a=Ak,$ where  $k'={2\pi F'/ H^{2} } $ and $F'$ is the second oscillation frequency, $\alpha $ and $\beta $ are ratios between oscillation amplitudes and frequencies respectively, we have the following equation for the thermodynamic potential density
$$\Omega ={\frac{1}{4\pi k^{2}} } \left[a\cos \left(\mu \right)+{\frac{\alpha a}{\beta}  } \cos \left(\beta \mu \right)+\mu ^{2} \right].  \eqno (6) $$                                                 
The new equation for the thermodynamic potential density gives a richer picture of the domain formation than that in the one-frequency case. In the two-frequency case, the instability of an electron gas does not occur at $a=1$.  

\section{Results and discussions}

      Condon domains were observed in aluminum by $\mu $SR spectroscopy \cite{b24} thereby validating results of the earlier work \cite{b25}, in which the effect of the phase transition on helicon waves was studied. We demonstrate here the thermodynamic potential density as a function of the dimensionless magnetization $\mu $ at different temperatures in aluminum in the two-frequency case (equations (3) and (6)).  According to the known theory of diamagnetic phase transitions at a $<$ 1 in the case under consideration at $a=0.3$ there is one minimum for $\mu  = 0$. The starting point of the appearance of two minima is $a = 0.8$ for $\alpha= 0.6$, $\beta  = 2.5$. Condon domains appear as a result of a second order phase transition\cite{b22}. One can obtain the same result by expansion of the thermodynamic potential density in powers of $\mu $.                                                                

 $$\mu =\sqrt{  \frac{  6\left[ a\left(1+\alpha \beta \right)-2\right]}{a\left(1+\alpha \beta ^{2} \right)}   }  \eqno (7) $$ 

It is seen from (7) that for $\alpha = 0.6$ and $\beta  = 2.5$ and $a = 0.8$ the value of $\mu $ is equal to zero. In Fig. 1 the thermodynamic potential density is shown as a function of $\mu $ at various values of $a$.  It is seen from Fig. 1 that there are two domains at $a = 4.0$, while at $a = 8.2$ four domains appear.  The two external minima are metastable, while the internal minima, corresponding to the two original domains, are stable. One can see from Fig.1 that a first-order phase transition takes place at $\alpha = 9.138$: all the four minima have the same depth. At larger values of a (for instance at $a=11.8$) the external domains are stable, while the internal domains are metastable. This means that a first-order phase transition occurs and metastable Condon domains are formed. 

      By using equation (3) for aluminum we construct the phase diagram temperature-magnetic field in the two-frequency case when $T_{D} =0.1K$ using $\varepsilon _{F} =3.76eV$ and $\eta _{c} ={m_{c}  \mathord{\left/{\vphantom{m_{c}  m_{e} =1.3}}\right.\kern-\nulldelimiterspace} m_{e} =1.3} $ \cite{b1}, where $m_{e} $ is the electron mass (Fig.2). At a = 0.8 we obtain the curve of second-order phase transitions.  As is seen from Fig.2, the phase transition occurs at $T=1.2K$ for $H=4.8T$ according to the measurements \cite{b24}. Starting from $a = 7.4$ the domain metastability region appears. At $a = 9.138$ we obtain the curve of first-order phase transitions. At $a = 11.95$ the four domains disappear. Condon domains in aluminum have been observed by measurements of magnetization oscillations of the relaxation rate of the muon precession polarization. According to the author's interpretation, the line doublet indicating the appearance of two Condon domains was not resolved by Fourier analysis due to low resolution of $\mu$SR spectrometer which was not able to detect the expected splitting of induction lines. We think that the absence of the doublet splitting may be due to the presence of numerous Condon domains taking place in the metastability range. The doubling of Condon domains accompanied by metastable phenomena is characteristic of the two-frequency case. It can be detected by improving the resolution of measurements of the $\mu $SR spectrum. Within a slight deviation from the center of magnetization oscillations, second-order and first-order diamagnetic phase transitions occur. 
\begin{figure}[h]
\centering  
\includegraphics[scale=0.3]{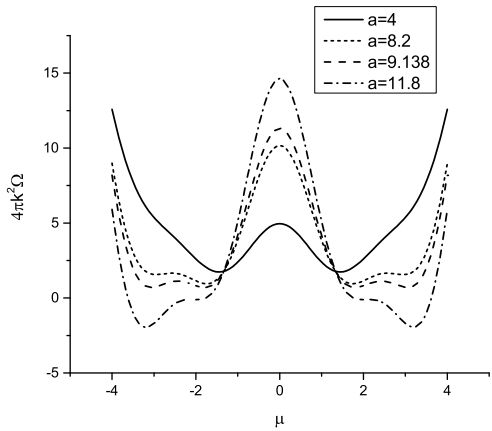} 
\caption{The dimensionless thermodynamic potential density $4\pi k^{2} \Omega $ as a function of the dimensionless magnetization $\mu $ for $\alpha = 0.6$, $\beta  = 2.5$: 1. $a = 4$ - two stable domains; 2. $a = 8$ - two metastable external domains and two stable internal domains; 3. $a = 9.138$ - four domains have the same depth. The first-order phase transition occurs; 4. $a = 11.8$ - two metastable internal domains and two stable external domains. }
\label{difmu2} 
\end{figure}

    The first direct observation of Condon domains was detected in silver, in which magnetization oscillations are due to belly and rosette oscillations \cite{b2}. The authors of Ref. \cite{b2} communicated that: ``During the temperature dependence studies it was noted that there was no magnetic induction splitting to two domains above $2.5K$ and that the existence of a splitting at $2.2K$ depended on whether the sample has been heated or cooled to that temperature, consistent with a supercooling effect''. The measured temperature hysteresis which is approximately equal to $0.3K$ should be related to the obtained first-order phase transition. As is seen from the phase diagram for aluminum, the temperature hysteresis is about $0.4K$ at $H = 4.8T$, whereas the observed temperature hysteresis in silver is $0.3K$, which is close to the calculated one. The appearance of an additional splitting of the magnetic induction lines in NMR and $\mu $SR experiments should be expected in the metastability range. The doubling of the number of Condon domains accompanied by metastable phenomena is characteristic of the two-frequency case. It can be detected by improving the resolution of NMR and $\mu $SR measurements. However, it may be observed in the narrow range of temperature and magnetic fields. By using the temperature-magnetic field phase diagrams one can find the metastability range and detect first-order phase transitions.  
\begin{figure}[h]
\centering  
\includegraphics[scale=0.3]{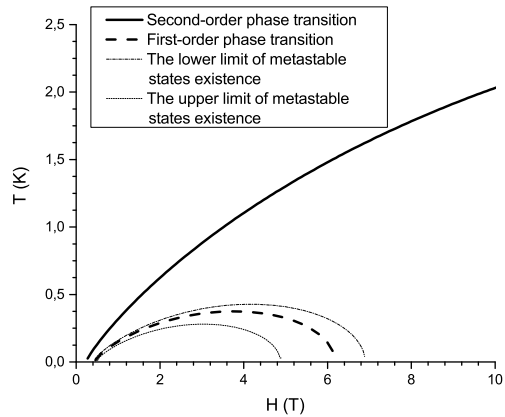} 
\caption{The phase diagram -- temperature-magnetic field -- in aluminum for $T_{D} =0.1K$. At $a = 0.8$ there is a curve of second-order phase transitions. Starting from $a = 7.4$ the domain metastability region appears. At $a = 9.138$ is a curve of first-order phase transitions. At $a = 11.95$ the four domains disappear and only two domains remain. }
\label{difmu1} 
\end{figure}

\section{Summary}

      Metastability effects in the formation of Condon non-spin magnetic domains have been considered. A possibility for the first-order phase transition occurrence in a three-dimensional electron gas has been studied in the case of two-frequency magnetization oscillations originating from two extremal cross sections of the Fermi surface. The appearance of two additional domains has been shown in the metastable region. The phase diagram exhibits the presence of second-order and first- order phase transitions leading to formation of Condon domains in the two-frequency case. 

\bibliography{metformcondom}

\end{document}